\documentclass[aps,prl,amssymb,showpacs,superscriptaddress,twocolumn,nofootinbib]{revtex4}
\usepackage{graphicx}

\newcommand{\Case}[2]{{\textstyle \frac{#1}{#2}}}
\newcommand{\lP}{\ell_{\mathrm P}}

\begin{document}
%
\preprint{IMSc/2004/07/29}

\title{Genericness of Big Bounce in isotropic loop quantum cosmology}

\author{Ghanashyam Date}
\email{shyam@imsc.res.in}
\affiliation{The Institute of Mathematical Sciences\\
CIT Campus, Chennai-600 113, INDIA.}

\author{Golam Mortuza Hossain}
\email{golam@imsc.res.in}
\affiliation{The Institute of Mathematical Sciences\\
CIT Campus, Chennai-600 113, INDIA.}

\begin{abstract}
The absence of isotropic singularity in loop quantum cosmology can be
understood in an effective classical description as the universe exhibiting
a Big Bounce. We show that with scalar matter field, the big bounce is
generic in the sense that it is independent of quantization ambiguities
and details of scalar field dynamics. The volume of the universe at the
bounce point is parametrized by a single parameter. It provides a
minimum length scale which serves as a cut-off for computations of
density perturbations thereby influencing their amplitudes.
\end{abstract}


\maketitle



It has been long expected that the existence of singularity in
the classical general relativity which has been shown to be quite
{\em generic} thanks to the {\em singularity theorems}, will be
removed when classical framework of gravity is extended to a quantum
framework of gravity. Despite of tremendous efforts made, unfortunately 
we still do not have a fully satisfactory theory of quantum gravity.
In the last couple of decades two strong contenders have emerged: the {\em 
string theory} approach \cite{StringRev} and the {\em loop quantum gravity} 
(LQG) approach \cite{LQGRev}. The issue of fate of classical
cosmological singularities has been addressed head-on within the LQG
approach. More precisely, the adaptation of LQG methods to cosmological
context, known as loop quantum cosmology (LQC) \cite{LQCRev}, has made 
impressive progress on the issue of singularities. It has been shown that the
isotropic models (flat and closed)\cite{Absence}, and more generally the 
diagonal Bianchi class A models \cite{HomCosmo,Spin}, are free of singularity. 
It is time now to explore further consequences of LQC corrected cosmologies.

Some of consequences of LQC corrected cosmologies have already been
noted. First, there is a natural mechanism for inflation
\cite{InflationB,InflationBV,GenericInflation} within the context of
isotropic models. Secondly, for the Bianchi IX model there is a
suppression of chaotic approach to singularity \cite{BianchiIX}.
Thirdly, there is indication of a bounce at the big crunch singularity
as well \cite{BounceClosed,Vereshchagin}. All of these have been explored within the
framework of an {\em effective Hamiltonian} which incorporates the most
significant non-perturbative corrections. These modifications stem from
the non-trivial definition of the inverse triad operator in LQC
\cite{InvScale,Spin} which ensure that the matter density, spin
connection components remain bounded as universe approaches zero volume.

The effective Hamiltonian is derived from the admissibility of a
continuum approximation \cite{Fundamental}. This comes about as a
requirement on the LQC quantization of the Hamiltonian constraint
operator which always leads to a {\em difference equation} for the
quantum wave function. For large volumes, these wave functions are
expected to vary slowly since quantum effects are small. This feature
allows one to use an interpolating wave function of a continuous
variable which satisfies a {\em differential equation}, the
`Wheeler--DeWitt equation'. Making a further WKB approximation leads one
to a Hamilton-Jacobi equation from which the effective Hamiltonian is
read-off. This is extrapolated to smaller (but not too small) volumes
by {\em replacing} the occurrence of inverse triad/volume factors by a
function coming from the definition of the inverse triad operator. The
validity of the effective Hamiltonian is limited by the validity of WKB
approximation and the validity of the continuum approximation.

Recently, we have extended the domain of validity of the continuum
approximation by exploiting the non-separable structure of the
Kinematical Hilbert space of LQC \cite{Bohr} which has infinitely many
solutions of the fundamental difference equation. Although each of these
may not be slowly varying at smaller volumes, one can choose linear
combinations to construct solutions which are slowly varying almost every 
where. This amounts to an ad-hoc restriction to a sub-class of solution. In
the absence of any other criteria to limit the infinity of solutions, such
as a physical inner product, this restriction is treated as exploratory.
The extraction of effective Hamiltonian then follows the same
method as before via a WKB approximation. The validity of the effective 
Hamiltonian is now limited only by the validity of WKB approximation i.e. 
to `classically accessible regions'. The effective Hamiltonian is derived 
in \cite{EffectiveHamiltonian} and is given by,
\begin{equation}
H^{\text{eff}} =  -~\frac{1}{\kappa}
\left[ \frac{B_+(p)}{4 p_0} K^2 + \eta \frac{A(p)}{2 p_0} \right]
 + W_{qg} + H_{m}
\label{EffHamiltonian}
\end{equation}
where $\kappa = 16\pi G$, 
$p_0 = \frac{1}{6}\gamma \lP^2 \mu_0$, $K$ is the extrinsic
curvature (conjugate variable of $p$),
$A(p)= |p+p_0|^{\frac{3}{2}} - |p-p_0|^{\frac{3}{2}}$, 
$\eta$ takes values $0,1$ for spatially flat, closed models
respectively, $B_+(p) = A(p+4p_0)+ A(p-4p_0)$, $\lP^2:=\kappa
\hbar$ and $ W_{qg} = \left(\frac{\lP^4}{288 \kappa p_0^3}\right)
\left\{B_+(p) - 2 A(p) \right\}$.
Apart from the modification of the coefficients of the gravitational
kinetic term and the spatial curvature term, the effective
Hamiltonian(\ref{EffHamiltonian}) differs from the classical
Hamiltonian by a non-trivial potential term referred to as {\em quantum
geometry potential} and denoted as $W_{qg}$. It is odd under the reversal
of orientation of the triad ($p \to - p$) and for $p > 0$, it is negative 
definite. The origin of this potential term is necessarily quantum 
gravitational as it explicitly involves $\lP$. For large volume this 
potential falls-off as $p^{-3/2}$ while for small volume it vanishes as
$p$. These small/large volume regions are delineated by the scale
$p_0$. 



For simplicity we consider a matter sector consisting of a single scalar 
field. Its classical Hamiltonian is LQC corrected in the usual manner 
\cite{InflationB,ScalarMatter}. It is shown in \cite{GenericInflation} that 
for small volume ($p \ll 2j p_0$), with non-perturbative corrections, the
scalar field effectively behaves like an {\em inflaton} field since the
effective equation of state variable, $\omega^{\text{eff}} \to - 1$ (or $-
\Case{4}{3}$ if the triad variable $p$ can get smaller than $p_0$). 
The matter Hamiltonian is related to the matter effective density
\cite{EffectiveHamiltonian,GenericInflation} as $H_m = \Case{3 p_0}{8}
\Case{a^4}{B_+} \rho^{\text{eff}}$ and the conservation equation implies 
that for a constant $\omega^{\text{eff}}$, the effective density goes as 
$\sim a^{-3(1 + \omega^{\text{eff}})}$. One can see that in either case of 
inflationary or super-inflationary regimes, the matter Hamiltonian always 
goes as $\sim p^{3/2}$.

\begin{figure}[htb]
\begin{center}
\includegraphics[width=8cm]{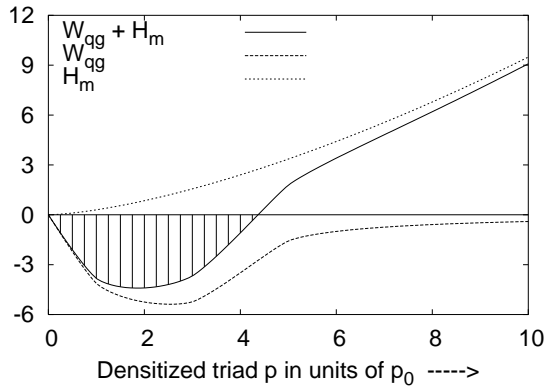}
\end{center}
\caption{For the small volume with non-perturbative corrections, the scalar
matter Hamiltonian along a trajectory is given by
$H_m \approx h p^{\frac{3}{2}}$. Thus non-perturbatively corrected scalar
matter Hamiltonian vanishes at $p=0$. If there were no quantum geometry 
potential term, $W_{qg}$, in the effective Hamiltonian then for the spatially
flat case ($\eta=0$) $p=0$ point would have been accessible through the 
evolution.  Although being a non-singular evolution it would have taken
infinite coordinate time to reach $p=0$ nevertheless there would have been
no {\em minimum} proper length for the given space-time.  But once we 
incorporate the effect of quantum geometry potential $W_{qg}$ then we
can see that the combined effect of $H_{m}$ and $W_{qg}$ will lead the 
extrinsic curvature to become zero at a non-zero value of $p$. Since for 
small $p$, $W_{qg} \sim -p$ and $H_{m} \sim p^{\frac{3}{2}}$ then there will
always be a region where $W_{qg}$ dominates over $H_{m}$. Naturally there
will always exist a classically inaccessible region leading to a generic
{\em Big Bounce}.  The shaded region in the figure represents the classically
forbidden region.} \label{bouncefig}
\end{figure}

Given the behaviors of the quantum geometry potential and the matter
Hamiltonian during a (super) inflationary region and their opposite signs,
it is clear that the quantum geometry potential will always dominate the 
matter Hamiltonian implying imaginary value for the extrinsic curvature i.e.
existence of classically in-accessible scale factors. The two necessarily
cancel each other at a finite, non-zero value of the scale factor. This would
be so even after including the contribution of the spatial curvature ($\eta$).
But this means that the extrinsic curvature vanishes at that value of the 
scale factor implying a bounce. Thus we see that a bounce is quite generic
and the minimum scale factor defines a new length scale $L_{\text{bounce}}$.
Below this scale, the effective classical picture fails. A graphical 
illustration of the existence of bounce can be seen in the figure 
(\ref{bouncefig}).

The bounce scale is obviously determined by the conditions
$H^{\text{eff}} = 0 = K$ with $H_m = h p^{3/2}$, $h$ being a
constant of proportionality. This is a transcendental equation in $p$
and the root(s) depend on the constant $h$. We expect the bounce value
to be {\em less} than $2j p_0$ (above which we are in the classical
regime). In geometrized units, $\kappa = 1 = c$, we will refer to all
lengths to the scale $\sqrt{p_0}$. Thus, putting $p := q p_0$ the region
of interest is $0 \ll q \ll 2j$. This could be further divided into (i)
$0 \ll q \ll 1$ and (ii) $1 \ll q \ll 2j$. Since the equation of state 
variable is a dimensionless function of the scale factor, it is in fact a 
function of $q$. The conservation equation can then be solved as:
$\rho^{\text{eff}}(q) = \bar{\rho}\ \text{exp} \{- \Case{3}{2}
\int_{2j}^q (1 + \omega^{\text{eff}}(q))\Case{dq}{q} \}$. The constant
$h$ is proportional to $\bar{\rho} = \rho^{\text{eff}}(2j)$. In summary,
the equations determining the (non-zero) bounce scale,
$q_{\text{bounce}}$, are,

\begin{eqnarray}
H_m(q) & = & 6 \frac{p_0^3}{B_+} q^2 \bar{\rho}\ \text{exp} 
\left\{-\Case{3}{2} \int_{2j}^q (1 + \omega^{\text{eff}}(q))\Case{dq}{q} 
\right\} \label{BounceEqnOne}\\
\parallel ~~~ & \nonumber \\
h \ (p_0 q)^{\Case{3}{2}} & = & \eta \Case{A(q)}{2 p_0} - 
\Case{\lP^4}{288 p_0^3}(B_+(q) - 2 A(q)) \label{BounceEqnTwo}
\end{eqnarray}
In the two regions (i) and (ii), the equations simplify. In the region (i) 
one has, $ A \to 3 p_0^{3/2} q, B_+ \to 3(\sqrt{5} - \sqrt{3}) p_0^{3/2} q, 
\omega^{\text{eff}}(q) \to - \Case{4}{3}$ and we get, 
\begin{eqnarray}
H_m(q) & \to & \left[\sqrt{\frac{2}{j}}\frac{\bar{\rho}}{(\sqrt{5} -
\sqrt{3})} \right] (p_0 q)^{\frac{3}{2}} \label{Ham1}\\
\sqrt{q}_{\text{bounce}} & = &  \frac{1}{2 p_0 h} \left[ 3 \eta + 
\frac{(2 - \sqrt{5} + \sqrt{3})}{48} \frac{\lP^4}{p_0^2} \right] .\label{Soln1}
\end{eqnarray}
In region (ii) one has $ A \to p_0^{3/2}(3 q^{1/2} - \Case{1}{8}
q^{-3/2}),  B_+ \to p_0^{3/2}(6 q^{1/2} - \Case{49}{4}
q^{-3/2}), \omega^{\text{eff}}(q) \to - 1$ and
we get,
\begin{equation}
H_m(q) ~ \to ~ \bar{\rho} (p_0 q)^{\Case{3}{2}}\ , \label{Ham2}
\end{equation}
and the $q_{\text{bounce}}$ is determined as a root of the cubic equation,
\begin{equation}
(2 p_0 \bar{\rho}) q^3 - 3 \eta q^2 - \frac{1}{12} 
\left(\Case{\lP^4}{p_0^2}\right) ~ = ~ 0\label{Eqn2}
\end{equation}
Note that $p_0 \bar{\rho}$ is dimensionless. It is easy to see that
there is exactly one real root of this equation and in fact, get a close
form expression for it.

In the region (i), $h \sim \bar{\rho}/\sqrt{j}$ and the $q_{\text{bounce}}
\sim (p_0 h)^{-2}$. The solution has explicit $j$ dependence. The inequality
for region (i), $q_{\text{bounce}} \ll 1$, implies that the effective density
at $p = 2j p_0$ (roughly where the inverse scale factor function attains its
maximum) must be larger than $\sqrt{j}$ times the Planck density $\sim 
\lP^{-2}$. Also the bounce scale will be smaller than the Planck scale.
This is indicative of the effective continuum model becoming a poor
approximation.

\begin{figure}[htb]
\begin{center}
\includegraphics[width=8cm,height=6cm]{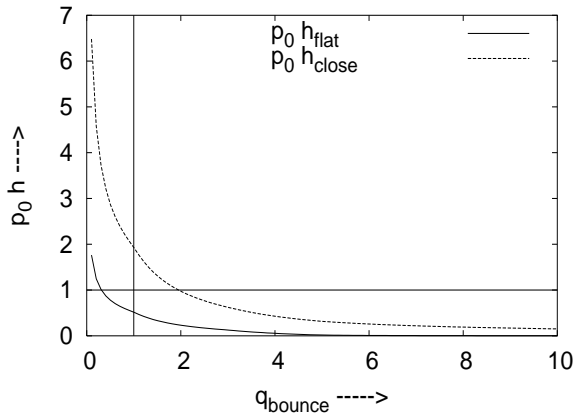}
\end{center}
\caption{The plot shows how the dimensionless parameter $hp_0$ varies as
a function of the bounce scale $q_{\text{bounce}}$ as determined by the
equations (\ref{BounceEqnOne}, \ref{BounceEqnTwo}). $p_0 = \Case{1}{6}\lP^2$
has been chosen for the plot.} \label{DensityCorr}
\end{figure}

In region (ii), $h = \bar{\rho}$. For the flat model ($\eta = 0$), 
$q_{\text{bounce}} \sim (p_0 \bar{\rho})^{-1/3}$. For the close model,
such a simple dependence does not occur. The inequalities, $ 1 \ll q \ll 2j$,
translate into a window for $p_0 \bar{\rho}$. Region (ii) bounce scale has 
{\em no explicit} dependence on the ambiguity parameter $j$, the
implicit dependence being subsumed in the value of $\bar{\rho}$ which
can be treated as a free parameter. The bounce scale is larger than the
Planck scale and the density $\bar{\rho}$ is smaller than the Planck
density. The relation between $h$ and the bounce scale is displayed in
figure (\ref{DensityCorr}).

Clearly, as $\hbar \to 0$, the scale $p_0 \to 0$ and so does the bounce
scale. It also vanishes as $\bar{\rho} \to \infty$. However in the
non-singular evolution implied by isotropic LQC (inverse scale factor
having a bounded spectrum) implies that there is a maximum energy
density attainable and the density $\bar{\rho}$ can be thought of as
this maximum energy density. Correspondingly there is a minimum scale
factor or minimum {\em proper} volume since in LQC the fiducial coordinate
volume is absorbed in the definition of the triad. For volumes smaller 
than the minimum volume, the WKB approximation fails and the effective 
classical picture cannot be trusted. The quantum geometry potential plays
a crucial role in this result. 

A remark about the physical justification for the approximations used is in
order. The results use the effective Hamiltonian picture which is based on  a
continuum approximation followed by the WKB approximation. The physical
justification thus hinges on the physical justification for these two
approximations.  The continuum approximation for the geometry is physically
expected to be a good approximation for length scales larger than the
discreteness scale set by $4p_0 \propto 4\mu_0$, the step size in the
fundamental difference equation. The WKB approximation is valid in a sub-domain
of the continuum approximation, determined by slow variation of amplitude and
phase.  Mathematically, the amplitude variation begins to get stronger around
$2p_0$ \cite{EffectiveHamiltonian} while the phase variation is stronger
at the turning point which determines the bounce scale. Thus, physically, the effective
Hamiltonian (including the quantum geometry potential) is trustworthy for the
bounce scale larger than $\sim p_0$. As shown by the figure 2, there is a range
of $\bar{\rho} < l_p^{-2}$ such that the bounce scale is consistent with the
physical domain of validity of the effective Hamiltonian. Note also that the
behavior of the matter Hamiltonian as $p^{\Case{3}{2}}$ is dependent on $p \ll
2j p_0$ and hence the bounce scale is also smaller than $2j p_0$.

We will discuss now a possible implication of the minimum proper length
on the inflationary cosmology.  The standard inflationary scenario is 
considered a successful {\em paradigm} not only because it can effectively
solve the traditional problems of standard classical cosmology, but also 
because it provides a natural mechanism of generating classical {\em seed} 
perturbations from quantum fluctuations. These seed perturbations are 
essential in a theory of large scale structure formation but there is no 
mechanism of generating the initial perturbation within the classical setup.
A quantum field living on an inflating background quite generically produces 
{\em scale-invariant} power spectrum of primordial density perturbations 
which is consistent with the current observations.

However, one major problem that plagues almost all potential driven 
inflationary models is that these models generically predict too much 
{\em amplitude} for density perturbation \cite{Brandenberger, Narlikar}. 
Considering the fluctuations of quantum scalar field on an inflating 
classical background, one can show that these models {\em naturally} predict
density perturbations at {\em horizon re-entry} to be $\frac{\delta
\rho}{\rho} \sim 1 - 10^2$. But CMB anisotropy measurements indicates 
$\frac{\delta \rho}{\rho} \sim 10^{-5}$. Thus it is very difficult to get
desired amplitude for density perturbation from the standard inflationary 
scenarios unless one introduces some {\em fine tuning} in inflaton potential
\cite{Guth}.

An interesting suggestion to get an acceptable amount of density perturbation
from inflationary scenario was made by Padmanabhan
\cite{Padmanabhan,PadmanabhanDetail}. The basic idea of the suggestion is 
that any proper theory of quantum gravity should incorporate a {\em zero-point}
proper length. This in turns damps the propagation of modes with proper 
wavelength smaller than the zero-point proper length. This mechanism reduces
the amplitude of density perturbation by an exponential damping factor.
The computations of \cite{PadmanabhanDetail} show that with the energy
density ($V_0$) attainable during inflation to be order of the Planck energy
density and the introduced cutoff ($L$) of the order of $(G\hbar/c^3)^{1/2}$,
one can indeed get the necessary amount of damping. In the picture
discussed above, we already have a correlation between $\bar{\rho}$ and
the bounce scale.

As discussed above, the effective model derived from semi-classical LQC
already shows the existence of a minimum proper length which can play
the role of the zero-point length. Furthermore, this scale is {\em not}
put in by hand but arises generically and naturally from the non-singularity
of the effective model \cite{EffectiveHamiltonian} and is correlated
self-consistently with the maximum attainable energy density whose
existence is also guaranteed in a non-singular evolution. It has no
explicit dependence on the quantization ambiguity parameter $j$.
With the genericness of (exponential) inflation shown in \cite{GenericInflation}
one can expects that the effective LQC model has the potential to produce
an acceptable primordial power spectrum as well as an acceptable amplitude
for density perturbation. A detailed analysis of primordial density 
fluctuations incorporating semi-classical LQC modifications is being
carried out and will be reported elsewhere \cite{DensityPert}.

Apart from the possible phenomenological implications of the existence
of a bounce, there are some theoretical implications as well. Within the
WKB approximation used in deriving the effective Hamiltonian, existence
of bounce corresponds to existence of classically in-accessible regions
(volumes). This can also be interpreted as limiting the domain of
validity of continuum geometry or the kinematical framework of general
relativity. Since the exact quantum wave functions do connect the two 
regions of the triad variable, there is also the possibility of tunneling
to and from the oppositely oriented universe ($p < 0$) through these
regions. Because of this, the bounce can be expected to be `fuzzy'. If and 
how the tunneling possibility between oppositely oriented universe affects 
`discrete symmetries' needs to be explored. 

Finally, the bounce result has been derived using genericness of inflationary
regime ($p \ll 2j p_0$). It is reasonable to assume that the maximum
energy density would be comparable or less than the Planck density. In
such a case the bounce scale will be greater than $p_0$. Thus, both the
results regarding genericness of bounce and genericness of inflation would
follow even if the underlying assumption of slowly varying wave
functions is valid only down to the bounce scale.

\begin{acknowledgments}
We are grateful to Martin Bojowald for helpful comments.
\end{acknowledgments}


\end{document}